\documentclass{article}
\usepackage{dcase2018,amsmath,graphicx,url,times,booktabs, tabularx}
\usepackage[usenames, dvipsnames]{color}

\newcommand{\mylabel}[2]{#2\def\@currentlabel{#2}\label{#1}}

\title{To bee or not to bee: Investigating machine learning approaches for \\beehive sound recognition}

\name{In\^{e}s Nolasco and Emmanouil Benetos\thanks{This work was supported by UK EPSRC grant EP/R01891X/1 and a UK RAEng Research Fellowship (RF/128).}}
\address{School of Electronic Engineering and Computer Science, Queen Mary University of London, UK\\ \texttt{\{i.dealmeidanolasco@,emmanouil.benetos@\}qmul.ac.uk}}
\begin{document}

\ninept
\maketitle

\begin{sloppy}

\begin{abstract}

In this work, we aim to explore the potential of machine learning methods to the problem of beehive sound recognition.
A major contribution of this work is the creation and release of annotations for a selection of beehive recordings. By experimenting with both support vector machines and convolutional neural networks, we explore important aspects to be considered in the development of beehive sound recognition systems using machine learning approaches.

\end{abstract}


\section{Introduction}
\label{sec:intro}

A significant part of computational sound scene analysis research involves the development of methods for automatic analysis of sounds in natural environments. This area of research has close links with the field of bioacoustics and has several applications, including automatic biodiversity assessment and automatic animal welfare monitoring \cite{Stowell18}. Within the context of \emph{computational bioacoustic scene analysis}, the development of technologies for automated beehive monitoring has the potential to revolutionise the beekeeping profession, with benefits including but not limited to a reduction of manual inspections, distant monitoring of bee populations, and by rapidly identifying phenomena related to the natural cycle of the beehive (e.g. queen missing, bee swarming).

In particular, sound plays a central role towards the development of such technologies for automated beehive monitoring. In \cite{Bencsik2011, Zacepins2016}, the authors give a thorough description of bee sounds and their characteristics. In short, the sound of a beehive is a mixture of the individual contributions of sounds produced by each bee of the colony. This mixture is perceived as a dense, continuous, low-frequency buzz. 

The first step towards the creation of audio-based beehive monitoring technologies is to create systems that are able to recognise bee sounds and discriminate them from other sounds that might be captured. These non-bee sounds will usually be related with the environment and events occurring in the hive's surroundings and can be as varied as urban sounds, animals, rain, or maintenance sounds. Thus, the aim of this work is to automatically detect sounds produced by bees, distinguishing them from external non-related sounds, given audio recordings captured inside beehives. One aspect that appears useful to differentiate between both classes is that the majority of non-beehive sounds can be of a short duration when compared with beehive sounds.

Related works in beehive sound analysis generally use heavy data pre-processing, hand-crafted features and domain knowledge to clean the recordings and come up with useful representations for beehive audio signals. 
In \cite{Ferrari2008}, the authors apply at a first stage a Butterworth filter with cut-off frequencies of 100 Hz and 2000 Hz in order to filter the acoustic signal and remove all sounds of frequencies expected not to be in the bee sound class. 
In \cite{Robles-guerrero1870}, besides the use of several filtering techniques, 
the authors propose the use of Mel-frequency cepstral coefficients (MFCCs) as features to represent beehive sounds, inspired by speech processing research.
The work of \cite{Amlathe2018} is directly relevant to this paper, since a classification is performed to clean the recordings from external sounds. This task is set up to distinguish between 3 classes: beehive sounds, environmental sounds and cricket sounds. However, denoising techniques and hand-crafted features are still applied, including Wavelet transforms and features such as MFCCs, chroma and spectral contrast.

Machine learning methods, and in particular deep learning methods, can decrease up to a point the amount of handcrafted features and domain knowledge which can be responsible for introducing bias and limiting the modelling capabilities of sound recognition methods. 
In \cite{Kiskin}, deep neural networks (DNNs) and convolutional neural networks (CNNs) are used to automatically detect the presence of mosquitoes in a noisy environment, although the proposed methodology disregards the long duration characteristics of mosquito sounds. 
The work of \cite{Grill2017} tackles the problem of detecting the presence of birds from audio as part of the 2017 Bird Audio Detection challenge\footnote{\url{http://machine-listening.eecs.qmul.ac.uk/bird-audio-detection-challenge/}}. The proposed method, \textit{Bulbul}, is a combination of deep learning methods also relying on data augmentation. Given that \textit{Bulbul} was the challenge submission that produced the best results, it became the baseline method for the DCASE 2018 Bird Audio Detection task\footnote{\url{http://dcase.community/challenge2018/task-bird-audio-detection}}. 
In the context of environmental sound scene analysis, it is shown in \cite{Li2017} that DNNs have good performance when compared to shallower methods such as Gaussian mixture models (GMMs). However the authors also stress that the use of temporal methods such as recurrent neural networks (RNNs) does not improve classification in this context, which they justify with the characteristic of environmental sounds as not having strong temporal dependencies and being rather non-predictive and random.

In this work, we aim to explore the potential of machine learning methods to the problem of beehive sound recognition, as a first step towards the creation of audio-based beehive monitoring systems.
A core problem when using supervised machine learning methods is the large amount of labelled data needed. A major contribution of this work is the creation and release of annotations for a selection of recordings from the Open Source Beehive project \cite{OSBH} and for a part of the NU-Hive project dataset \cite{StefaniaCecchi1AlessandroTerenzi1SimoneOrcioni1PaolaRiolo2SaraRuschioni22018}.
The annotated data is used in experiments using support vector machines (SVMs) and a CNN-based approach by adapting the \textit{Bulbul} implementation \cite{Grill2017}. The results presented are indicative of the important aspects to be considered in the development of machine learning-based beehive sound recognition systems.

The outline of the paper is as follows. 
In Section 2 we describe the data and the annotation procedure. Section 3 describes the methods applied; Section 4 presents the experiments performed, the evaluation metrics, and results. Finally, Section 5 concludes the paper and provides directions for future research.

\section{Data annotation} \label{sec:format}

The main issue of posing the problem of automatic recognition of beehive sounds as a classification problem is the need for annotated data. In this case we need examples of pure beehive sounds and examples of external sounds as they occur in the recordings made inside the beehives, so that the methods can learn their characteristics and map them to the corresponding labels. Given the lack of labelled data for this task, a major effort of developing such a dataset is undertaken here.
%
The resulting dataset is based on a selected set of recordings acquired in the context of two projects: the Open Source Beehive (OSBH) project \cite{OSBH} and the NU-Hive project \cite{StefaniaCecchi1AlessandroTerenzi1SimoneOrcioni1PaolaRiolo2SaraRuschioni22018}. 
The main goal of both projects is to develop beehive monitoring systems capable of identifying and predicting certain events and states of the hive that are of interest to beekeepers. Among many different variables that can be measured and that help the recognition of different states of the hive, the analysis and use of the sound the bees produce is a big focus for both projects.

The recordings from the OSBH project \cite{OSBH} were acquired through a citizen science initiative which asked members of the general public to record the sound from their beehives together with the registering of the hive state at the moment. Because of the amateur and collaborative nature of this project, the recordings from the OSBH project present great diversity due to the very different conditions in which the signals were acquired: different recording devices used, different environments where the hives were placed, and even different position for the microphones inside the hive. This variety of settings makes this dataset a very interesting tool to help evaluate and challenge the methods developed.

The NU-Hive project \cite{StefaniaCecchi1AlessandroTerenzi1SimoneOrcioni1PaolaRiolo2SaraRuschioni22018} is a comprehensive effort of data acquisition, concerning not only sound, but a vast amount of variables that will allow the study of bee behaviours. Contrary to the OSBH project recordings, the recordings from the NU-Hive project are from a much more controlled and homogeneous environment. Here the occurring external sounds are mainly traffic, honks and birds.

The annotation procedure consists in listening the selected recordings and marking the onset and offset of every sound that could not be recognised as a beehive sound. The recognition of external sounds is based primarily on the perceived heard sounds, but a visual aid is also used by visualising the log-mel-frequency spectrum of the signal.
All the above are functionalities offered by Sonic Visualiser\footnote{\url{http://sonicvisualiser.org/}}, which was used by two volunteers that are neither bee-specialists nor specially trained in sound annotation tasks.
By marking these pairs of instances corresponding to the beginning and end of external sound periods, we are able to get the whole recording labelled into \textit{Bee} and \textit{noBee} intervals.
The \textit{noBee} intervals refer to periods where an external sound can be perceived (superimposed to the bee sounds).
\begin{figure}[t]
  \centering
  \centerline{\includegraphics[width=.8\columnwidth]{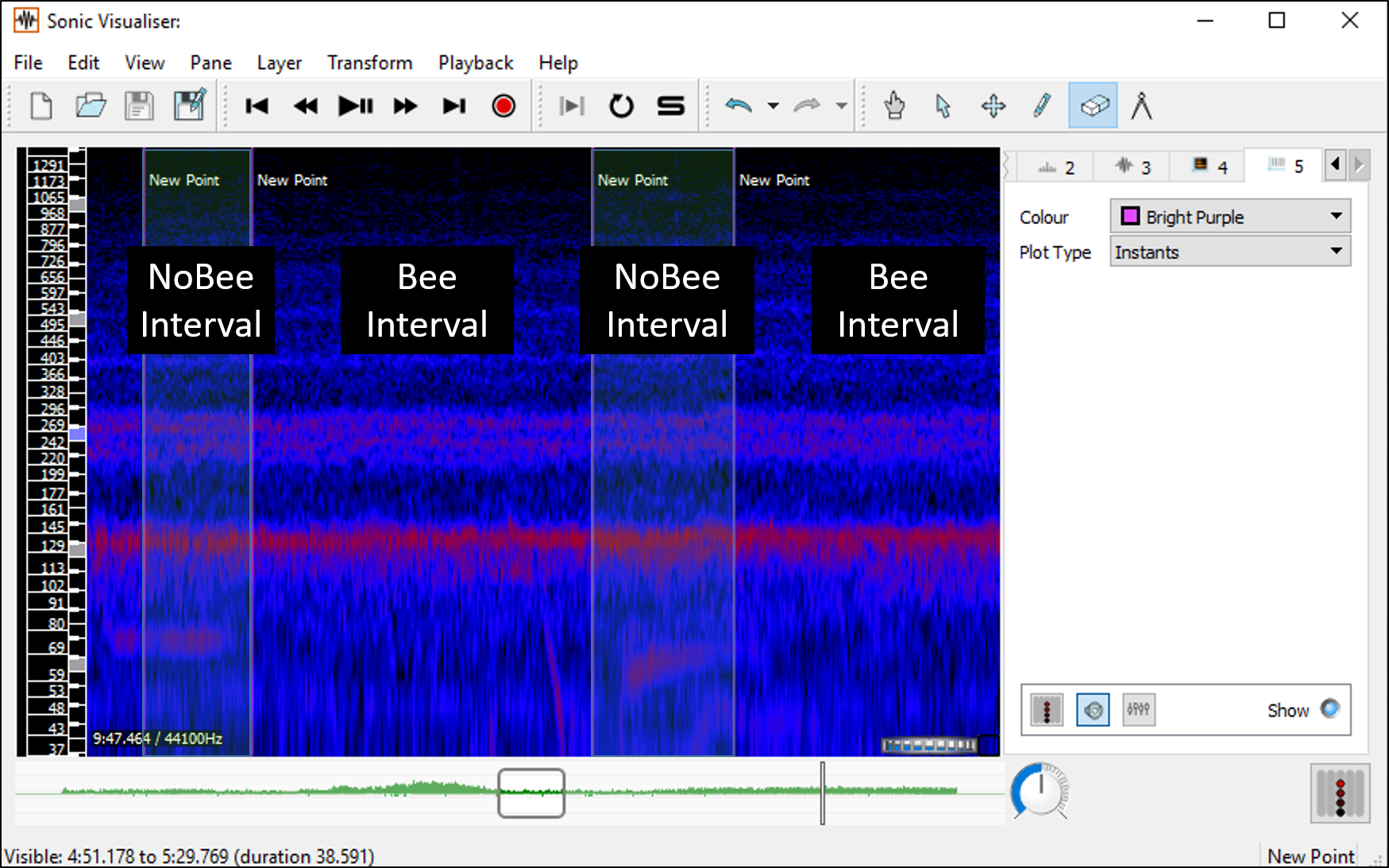}}
  \caption{Example of the annotation procedure for one audio file.}
  \label{fig:annotationExample}
\end{figure}
An example of this process is shown in Fig. \ref{fig:annotationExample}.



The whole annotated dataset consists of 78 recordings of varying lengths which make up for a total duration of approximately 12 hours of which 25\% is annotated as noBee events.  
About 60\% of the recordings are from the NU-Hive dataset and represent 2 hives, the remaining are recordings from the OSBH dataset and 6 different hives. The recorded hives are from 3 regions: North America, Australia and Europe. The annotated dataset\footnote{\url{https://zenodo.org/record/1321278#.W2XswdJKjIU}} and auxiliary Python code\footnote{\url{https://github.com/madzimia/Audio_based_identification_beehive_states}} are publicly available.

\section{Methods}
\label{sec:methods}

\subsection{Preprocessing}
 
The audio recordings are processed at a 22050~Hz sample rate, 
and are segmented in blocks of predefined lengths. Segments smaller than the defined block length have their length normalised by repeating the audio signal until the block length is reached.
For each block a label is assigned based on the existing annotations. A label \textit{Bee} is assigned if the entirety of the segment does not contain nor overlap any external sound interval. Similarly, the label \textit{noBee} is assigned if at least a part of the segment contains an external sound event. 
Finally, the training data is artificially balanced by randomly duplicating segments of the class less represented.

In order to evaluate the impact of the length of external sounds, 
we explore different threshold values ($\Theta$) for the minimum duration of external sounds to be included in the annotations. 


\subsection{SVM classifier}

We first create a system for beehive sound recognition using a support vector machine (SVM) classifier.
In order to gain insight on which features, normalisation strategies and other classifier parameters are promising to use in this problem, we explore a set of combinations of the three on the SVM classifier, detailed in Section \ref{sec:SVMexperiments}.
Two types of features are extracted for use with the SVM: 20 Mel-frequency cepstral coefficients (MFCCs) and Mel spectra \cite{chapter4_book}, the latter with 80 and 64 number of bands. The spectra are computed with a window size of 2048 samples and hop size of 512 samples.

\subsection{CNN classifier}


For the deep learning approach we explore the application of the \textit{Bulbul} CNN implementation \cite{Grill2017} as modified for the DCASE 2018 Bird Audio Detection task. 
The choice of this implementation for a first experiment using a deep learning approach is due to both its promising results achieved in the Bird Audio Detection Challenge, but also because the original problem for which the \textit{Bulbul} system was developed poses similar challenges as the ones we face. 

In this implementation, Mel spectra with 80 bands are computed using a window size of 1024 samples and a hop size of 315 samples. Additionally, these spectra are normalised by subtracting their mean over time.
%
%
The network consists of four convolution layers (two layers of 16 filters of size $3\times3$ and two layers of 16 filters of size $3\times1$) with pooling, followed by three dense layers (256 units, 32 units and 1 unit). 
All layers use a leaky rectifier as activation function with the exception of the output layer which uses the sigmoid function.



Data augmentation is also employed, which includes shifting the training examples periodically in time, and applying random pitch shifting of up to 1 mel band. 
Dropout of 50\% is applied to the last three layers during training.

\section{Evaluation}
\label{sec:Results}

\subsection{Experimental setup}
Given the diversity of the data available we are interested in evaluating how well the classifiers are able to generalise to different data. Thus, besides random splitting between train and test sets, we implement a ``hive-independent'' splitting scheme. This means having training samples belonging only to certain hives, and testing using samples from other, unseen hives.

For both schemes a test size of 5\% is used (5\% of the total number of segments in the case of the random split scheme or 5\% of the number of hives in the hive-independent splitting scheme). When applying the SVM classifier, all remaining data is used in a single training set. For the \textit{bulbul} implementation, in order to mimic the original cross validation scheme, where a model is trained in each set and validated on the others,
the remaining data (95\%) is further split in half between two sets.


The training of the \textit{Bulbul} network is done by stochastic gradient descent optimisation  on a mini-batch of 20 input samples of size 1000 frames by 
80 Mel-frequencies (receptive field), and through 100 epochs.
The 
training
samples are 
organised in 
two
sets, 
and
the resulting two trained models are ensembled to generate the predictions in the test set.
The prediction for a single sample is obtained by averaging the network output predictions of the non-overlapping 1000 frame excerpts that constitute the whole input sample.



\subsection{Evaluation Metrics}
 
The results of each experiment are evaluated using the area under the curve score (AUC) \cite{chapter6_book}. Each experiment is run three times following the same setup and parameters, and we report the results on each run and the average of the three. The results on the training set are also reported. 

\subsection{SVM Experiments} \label{sec:SVMexperiments}

As mentioned in Section \ref{sec:methods}, in this approach a combination of the below parameters is evaluated: 
\begin{description}
\item[SVM kernels:] RBF, linear, and 3\textsuperscript{rd} order polynomial.



\item[Features:]
$\mu$ and $\sigma$ of: 20 MFCCs, the $\Delta$ of 20 MFCCs and of the $\Delta\Delta$ of 20 MFCCs;
$\mu$ and $\sigma$ of: Mel-spectra and $\Delta$ of Mel-spectra with 64 or 80 bands; $\mu$ and $\sigma$ of: log Mel-spectra and $\Delta$ of log Mel-spectra with 64 or 80 bands; 


\item[Normalisation strategies:] no normalisation, normalisation by maximum value per recording, by maximum value in dataset, z-score normalisation at recording level, and z-score normalisation at dataset level.
\item[Segment size ($S$):] 30 seconds and 60 seconds.

\item[Threshold $\Theta$:] 0 seconds and 5 seconds.
\item[Split modes:] Hive-independent and Random split
\end{description}




Combining these parameters and evaluating the results of each combination leads us to define the optimal set of parameters (C*).
In order to thoroughly evaluate the classifier, experiments using C* are compared against specific parameter changes: (a) different value of threshold $\Theta$; (b) different segment size S; (c) Hive-independent split of the data to determine the generalisation capability to unseen hives; (d) Unbalanced dataset to determine the robustness of the classifier regarding unbalanced classes.

\subsection{CNN Experiments} \label{CNN_experiments}
Where possible, parallel experiments to the SVM approach are set up here.
As baseline parameters (B*), we use the following:
\begin{description}
\item[Features:] 80 Mel-band spectra
\item[Receptive field:] 1000 frames
\item[Number of training epochs:] 100
\item[Batch size:] 20
\end{description}

Experiments with changes to these parameters are: (a) different values of $\Theta$, to determine if the classifier can learn to reject only external sounds with long durations; (b) different values of segment size $S$; (c) Hive-independent split of data, to determine the generalisation capability of the classifier to unseen hives; (d) unbalanced dataset, to determine how the classifier can cope with this aspect; (e) larger receptive fields, to determine if the classifier can exploit the larger context of the input samples.


\subsection{SVM Results}


The resulting average AUC scores for the test and training set of the 3 runs of each experiment are shown in Fig.~\ref{fig:SVM_results}.
%
%
%
%
From the 1\textsuperscript{st} experiment we infer that the highest average AUC score in test sets is achieved when we use the following combination of parameters (C*): features as the $\mu$ and $\sigma$ of the value, the $\Delta$ and the $\Delta\Delta$ of 20 MFCCs, not considering the first coefficient; $S$ of 60 seconds, $\Theta$ of 5 seconds and not using any of the normalisation strategies defined. 


Fig.~\ref{fig:SVM_results} \textbf{[$\Theta$: 0sec]} shows the AUC results for the experiment using the C* parameters but changing $\Theta$ from 5 to 0 seconds. 
These show primarily that the classifier is not performing in a consistent way, which may 
indicate a strong dependency on the individual instances in which it is being tested and trained. Also the larger difference between the scores in the train and test sets indicate overfitting to the training examples.
Using the smallest value for $\Theta$ means that we provide to the classifier samples from which their label is defined based on what can be very short duration events. It is therefore expected that the classifier struggles to distinguish the classes. 

By running the classifier with C* parameters but with segment size changed from 60 to 30 seconds (Fig.~\ref{fig:SVM_results} \textbf{[$S$: 30sec]}), we can observe a decrease in both AUC in the train and test sets. These results affirm 
the idea that, given the long-term aspect of the beehive sounds, if we provide more context to the classifier, it will be better at distinguishing between the two classes of sounds.

In Fig.~\ref{fig:SVM_results} \textbf{[Hive-independent split]}, the classifier is run on 3 sets of data split using the hive-independent splitting scheme.
The results clearly show the inability of the classifier to generalise to unseen hives.

Fig.~\ref{fig:SVM_results} \textbf{[Unbalanced train-set]} shows the results of running the classifier in the same sets as experiment C*, but not replicating samples to artificially balance the sets.
Comparing the two, they are almost identical which makes sense for SVMs since when data balancing
is performed by simple data  duplication, the new points are all in locations where data points
already existed, therefore these do not influence the decision boundary found by the SVM.

\subsection{CNN Results}


\begin{figure}[t]
  \centering
  \centerline{\includegraphics[width=\columnwidth]{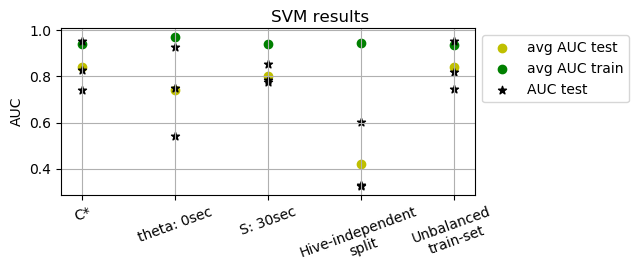}}
  \caption{SVM results on the test set for each of the 3 runs ($\star$), using the AUC score. The \textcolor{green}{\textbullet} and \textcolor{yellow}{\textbullet} represent the average AUC score of the 3 runs in both train and test sets respectively.}
  \label{fig:SVM_results}
\end{figure}

The resulting average AUC scores for the test and training sets for the 3 runs of each experiment are shown in Fig.~\ref{fig:bulbul_results}.
%
The first experiment determined that the best average AUC in the test sets of the 3 runs is achieved when we use the baseline parameters defined in \ref{CNN_experiments} plus the following parameters:
\textit{S} of 60 seconds and $\Theta$ of 0 seconds. The best results are shown in Fig.~\ref{fig:bulbul_results} \textbf{[B*]}. 

Regarding the values of $\Theta$, Fig.~\ref{fig:bulbul_results} \textbf{[$\Theta$: 5sec]} shows that using a larger $\Theta$ is detrimental to performance. This may be explained by the fact that the \textit{Bulbul} system was specifically designed for the detection of bird sounds, which are mainly short duration events, and thus struggles to identify longer events like traffic and rain sounds.

\begin{figure}[t]
  \centering
  \centerline{\includegraphics[width=\columnwidth]{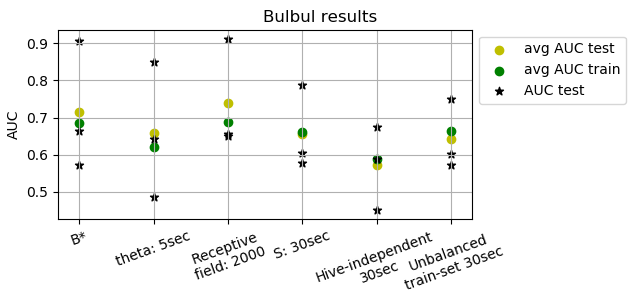}}
  \caption{Results for the \textit{Bulbul} CNN using the AUC score, for each of the 3 runs ($\star$). The \textcolor{green}{\textbullet} and \textcolor{yellow}{\textbullet} represent the average AUC score of the 3 runs in both train and test sets respectively.}
  \label{fig:bulbul_results}
\end{figure}

The experiment to evaluate if providing more context to the network improves performance is done by changing the receptive field from 1000 ($\sim$14 seconds) to 2000 ($\sim$30 seconds). 
In Fig.~\ref{fig:bulbul_results} \textbf{[Receptive field: 2000]}, the results show that indeed more context is particularly useful in the context of this problem. This is also consistent with the results from the SVM approach.

The role of \textit{S} in the CNN approach is different from the SVM one. Here, a larger segment size does not imply that larger samples with more context are given to the classifier, since this is controlled by the receptive field of the network. 
However, given that prediction is done for a whole segment by averaging the predictions for each frame, using larger segments leads to introducing more context.
Confirming the results regarding the need for more context, Fig.~\ref{fig:bulbul_results} \textbf{[S: 30sec]} shows that using a smaller segment size results in slightly worse predictions than using a larger segment size (\textit{S}: 60 seconds, shown in Fig.~\ref{fig:bulbul_results} \textbf{[B*]}).

Fig.~\ref{fig:bulbul_results} \textbf{[Hive-independent 30sec]} shows the results when using a hive-independent splitting scheme in a 30 second segment size data. Comparing this with the results in Fig.~\ref{fig:bulbul_results} \textbf{[S: 30sec]}, the lack of generalisation capacity to unseen hives is also evident here, although, compared with the SVM approach, the results seem to be slightly better and less overfitting occurs which may indicate better generalisation capabilities for the CNN.

Fig.~\ref{fig:bulbul_results} \textbf{[Unbalanced train-set 30sec]} shows the results of not doing data balancing on the 30 second segment data. When comparing with Fig.~\ref{fig:bulbul_results} \textbf{[S: 30sec]}, the results indicate that data balancing should be considered when training this CNN.


\section{Conclusions}
In this work we allocate a major effort for the creation of an annotated dataset for beehive sound recognition where machine learning approaches can be used. However, the annotation procedure can be improved for future additions to this dataset:
%
ideally annotations should be performed by specialists which label overlapping sets of data so that the annotations are subject to peer validation. 
Finally the main critique to the annotations could be that they are the most important source of human bias introduced in this work.

Although the scores achieved by the CNN implementation fail to achieve the level of the SVM approach, results are indicative of the important aspects to be considered when developing neural networks to tackle this unique problem. Mainly, the importance of providing samples with large context, the amount of training data, and finally due to the incapacity of both approaches to generalise to different hives, the one constraint would be to train systems in the same hives where they are going to be used.
We consider that this work can be a first step in a pipeline of beehive monitoring systems, which we think 
will have an important role in the future of bee keeping. 
Finally, we expect that this work and the release of the annotated dataset to further motivate research in this topic, and more broadly in the intersection of machine learning and bioacoustics.

\section{ACKNOWLEDGEMENT}
\label{sec:ack}
We would like to thank the authors of the NU-Hive project for creating such complete dataset and making it available for us to work with. 
Also a special thanks to Ermelinda Almeida for her effort and dedication on annotating the data.


\bibliographystyle{IEEEtran}
\bibliography{refs}

\begin{thebibliography}{10}
\providecommand{\url}[1]{#1}
\def\UrlFont{\rmfamily}
\providecommand{\newblock}{\relax}
\providecommand{\bibinfo}[2]{#2}
\providecommand\BIBentrySTDinterwordspacing{\spaceskip=0pt\relax}
\providecommand\BIBentryALTinterwordstretchfactor{4}
\providecommand\BIBentryALTinterwordspacing{\spaceskip=\fontdimen2\font plus
\BIBentryALTinterwordstretchfactor\fontdimen3\font minus
  \fontdimen4\font\relax}
\providecommand\BIBforeignlanguage[2]{{%
\expandafter\ifx\csname l@#1\endcsname\relax
\typeout{** WARNING: IEEEtran.bst: No hyphenation pattern has been}%
\typeout{** loaded for the language `#1'. Using the pattern for}%
\typeout{** the default language instead.}%
\else
\language=\csname l@#1\endcsname
\fi
#2}}

\bibitem{Stowell18}
D.~Stowell, ``Computational bioacoustic scene analysis,'' in
  \emph{Computational Analysis of Sound Scenes and Events}, T.~Virtanen, M.~D.
  Plumbley, and D.~P.~W. Ellis, Eds.\hskip 1em plus 0.5em minus 0.4em\relax
  Springer, 2018, pp. 303--333.

\bibitem{Bencsik2011}
M.~Bencsik, J.~Bencsik, M.~Baxter, A.~Lucian, J.~Romieu, and M.~Millet,
  ``{Identification of the honey bee swarming process by analysing the time
  course of hive vibrations},'' \emph{Computers and Electronics in
  Agriculture}, vol.~76, no.~1, pp. 44--50, 2011.

\bibitem{Zacepins2016}
A.~Zacepins, A.~Kviesis, and E.~Stalidzans, ``Remote detection of the swarming
  of honey bee colonies by single-point temperature monitoring,''
  \emph{Biosystems Engineering}, vol. 148, pp. 76--80, 2016.

\bibitem{Ferrari2008}
S.~Ferrari, M.~Silva, M.~Guarino, and D.~Berckmans, ``{Monitoring of swarming
  sounds in bee hives for early detection of the swarming period},''
  \emph{Computers and Electronics in Agriculture}, vol.~64, no.~1, pp. 72--77,
  2008.

\bibitem{Robles-guerrero1870}
A.~Robles-Guerrero and T.~Saucedo-Anaya, ``Frequency analysis of honey bee buzz
  for automatic recognition of health status: A preliminary study,''
  \emph{Research in Computing Science}, vol. 142, no. 2017, pp. 89--98, 1870.

\bibitem{Amlathe2018}
P.~Amlathe, ``Standard machine learning techniques in audio beehive monitoring:
  Classification of audio samples with logistic regression, {K}-nearest
  neighbor, random forest and support vector machine,'' Master's thesis, Utah
  State University, 2018.

\bibitem{Kiskin}
I.~Kiskin, P.~Bernardo, T.~Windebank, D.~Zilli, and M.~L. May, ``Mosquito
  detection with neural networks: The buzz of deep learning,'' \emph{ArXiv
  e-prints}, pp. 1--16, 2017, arXiv:1705.05180v1.

\bibitem{Grill2017}
T.~Grill and J.~Schl{\"{u}}ter, ``Two convolutional neural networks for bird
  detection in audio signals,'' in \emph{25th European Signal Processing
  Conference (EUSIPCO)}, 2017, pp. 1764--1768.

\bibitem{Li2017}
J.~Li, W.~Dai, F.~Metze, S.~Qu, and S.~Das, ``A comparison of deep learning
  methods for environmental sound detection,'' in \emph{IEEE International
  Conference on Acoustics, Speech and Signal Processing (ICASSP)}, Mar. 2017,
  pp. 126--130.

\bibitem{OSBH}
``{Open Source Beehives Project},'' \url{https://www.osbeehives.com/}.

\bibitem{StefaniaCecchi1AlessandroTerenzi1SimoneOrcioni1PaolaRiolo2SaraRuschioni22018}
S.~Cecchi, A.~Terenzi, S.~Orcioni, P.~Riolo, S.~Ruschioni, and N.~Isidoro, ``A
  preliminary study of sounds emitted by honey bees in a beehive,'' in
  \emph{Audio Engineering Society Convention 144}, 2018.

\bibitem{chapter4_book}
R.~Serizel, V.~Bisot, S.~Essid, and G.~Richard, ``Acoustic features for
  environmental sound analysis,'' in \emph{Computational Analysis of Sound
  Scenes and Events}, T.~Virtanen, M.~D. Plumbley, and D.~P.~W. Ellis,
  Eds.\hskip 1em plus 0.5em minus 0.4em\relax Springer, 2018, pp. 13--40.

\bibitem{chapter6_book}
A.~Mesaros, T.~Heittola, and D.Ellis, ``Datasets and evaluation,'' in
  \emph{Computational Analysis of Sound Scenes and Events}, T.~Virtanen, M.~D.
  Plumbley, and D.~P.~W. Ellis, Eds.\hskip 1em plus 0.5em minus 0.4em\relax
  Springer, 2018, pp. 13--40.

\end{thebibliography}
%
%
%
%
%
%
%
%
%

\end{sloppy}
\end{document}